Polarization superposition of room-temperature polariton condensation

Yuta Moriyama[1], Takaya Inukai[1], Tsukasa Hirao[1], Yusuke Ueda[1], Shun Takahashi[1], and Kenichi Yamashita[1*],

[1]*Faculty of Electrical Engineering and Electronics, Kyoto Institute of Technology, Matsugasaki, Sakyo-ku, Kyoto 606-8585, Japan*

*e-mail: yamasita@kit.ac.jp

A methodology for forming a qubit state is essential for quantum applications of room temperature polaritons. While polarization degree of freedom is expected as a possible means for this purpose, the coupling of linearly polarized polariton condensed states has been still a challenging issue. In this study, we show a polarization superposition of a polariton condensed states in an all-inorganic perovskite microcavity at room temperature. We realized the energy resonance of the two orthogonally polarized polariton modes with the same number of antinodes by exploiting the blue shift of the polariton condensed state. The polarization coupling between the condensed states results in a polarization switching in the polariton lasing emission. The orthorhombic crystal structure of the perovskite active layer and/or a slight off-axis orientation of the perovskite crystal axis from the normal direction of microcavity plane enable the interaction between the two orthogonally polarized states. These observations demonstrate a great promise of polariton as a room temperature qubit technology.

Introduction

Cavity polariton is a quasiparticle formed by the strong coupling between a photon and an exciton in an optically active microcavity.[1,2] As this bosonic quasiparticle possesses a small effective mass inherited from the photonic nature and a nonlinearity derived from the excitonic nature, the ensemble of cavity polariton particles reaches a threshold density for the Bose-Einstein condensation at relatively high temperature. This phenomenon is known as polariton condensation (or polariton lasing).[3,4] Furthermore, the room-temperature (RT) polariton condensation has also been revealed in several material systems, including widegap inorganic semiconductors,[5,6] organic semiconductors,[7–10] transition-metal dichalcogenides,[11,12] and lead-halide perovskites.[13–15] To utilize the RT polaritons for future quantum technologies, a methodology for forming a qubit state that can be readily created and manipulated should be established.[16,17] Polarization degree of freedom is a promising candidate as a means of forming the polaritonic qubit states, in analogous with the photonic quantum computing systems.[18] In fact, anisotropic optical features of polariton modes have been widely discussed in organic semiconductors[19–22] and perovskites.[23,24] However, a coupling of linearly polarized polariton condensed states has not been realized yet in RT polariton systems. It is a challenging issue to develop a strategy, in which an energy resonance and non-zero interaction between the two orthogonally polarized states are simultaneously obtained.

In this study, we demonstrate a polarization superposition of RT polariton condensed states. In an all-inorganic perovskite microcavity, we realize an energy resonance at the Γ point of the two orthogonally polarized modes by exploiting the blue shift of a polariton condensed state. The orthorhombic crystal structure of the perovskite active layer and/or a slight off-axis orientation of the perovskite crystal axis from the microcavity normal enable an interaction between the two polarized modes. Systematic photoluminescence (PL) results show that by a non-resonant optical pulse pumping, a polarization-hybridized condensed mode can be initialized to one of the polarization directions. As a result of the polarization hybridization of polariton condensed states, furthermore, we observed a polarization switching phenomenon in the polariton lasing operation. These observations demonstrate a great promise of polariton condensed states as a RT qubit technology.

Results and Discussion

In order to create polarization-coupled polariton condensed states, we employ an all-inorganic lead-halide perovskite, $CsPbBr_3$, as the anisotropic active medium in a microcavity. This compound has been demonstrated to show a high photoluminescence (PL) quantum yield and a large exciton binding energy, leading to a RT stability of exciton. Previous studies have experimentally demonstrated the RT polariton condensation in $CsPbBr_3$ microcavities.[13–15] A possibility of condensation in polaritonic Bardeen-Cooper-Schrieffer phase has been proposed for $CsPbBr_3$ microcavity at high excitation densities.[23,25]

Meanwhile, anisotropic polaritonic features also gather attention [see Fig. 1(a)]. Polarization anisotropy of $CsPbBr_3$ crystal, which shows the orthorhombic structure, breaks a rotation symmetry of the microcavity system, inducing a 'X-Y' polarization splitting in lower polariton branches (LPBs) at the Γ point [see Fig. 1(b)].[23,24] The LPB splits also into transverse-electric (TE) and transverse-magnetic (TM) modes at large wavevector regions because of quasi optical spin-orbit-coupling [see Fig. 1(b)].[23,26,27] These dispersion characteristics lead to polarization-dependent polariton lasing phenomena: Spencer et al. have reported that the split LPB modes show an independent polariton lasing at the energy position of each polarization mode.[23] Li et al. have reported in a study of liquid crystal-filled $CsPbBr_3$ microcavity that circularly polarized polariton lasing occurs from LPB modes split in in-plane wavenumber direction by optical Rashba-Dresselhaus coupling.[27] However, a coupling between linearly polarized condensed states has not been reported. Toward this purpose, we need to meet two requirements as follows: (i) a fine tuning of X-Y energy resonance and (ii) non-zero interaction between the X- and Y-polarized states. The X-Y energy resonance must be arranged for the LPB modes having the same number of antinodes as the spatial distribution of wavefunctions along the vertical direction should be the same for the two polarizations. In $CsPbBr_3$ microcavity, however, a fine tuning of the X- and Y-polarized modes with the same antinode number is impossible because they are originally detuned due to the X-Y splitting.

To overcome this problem, we exploit the blue shift of the polariton condensed state that emerges at the above-threshold density.

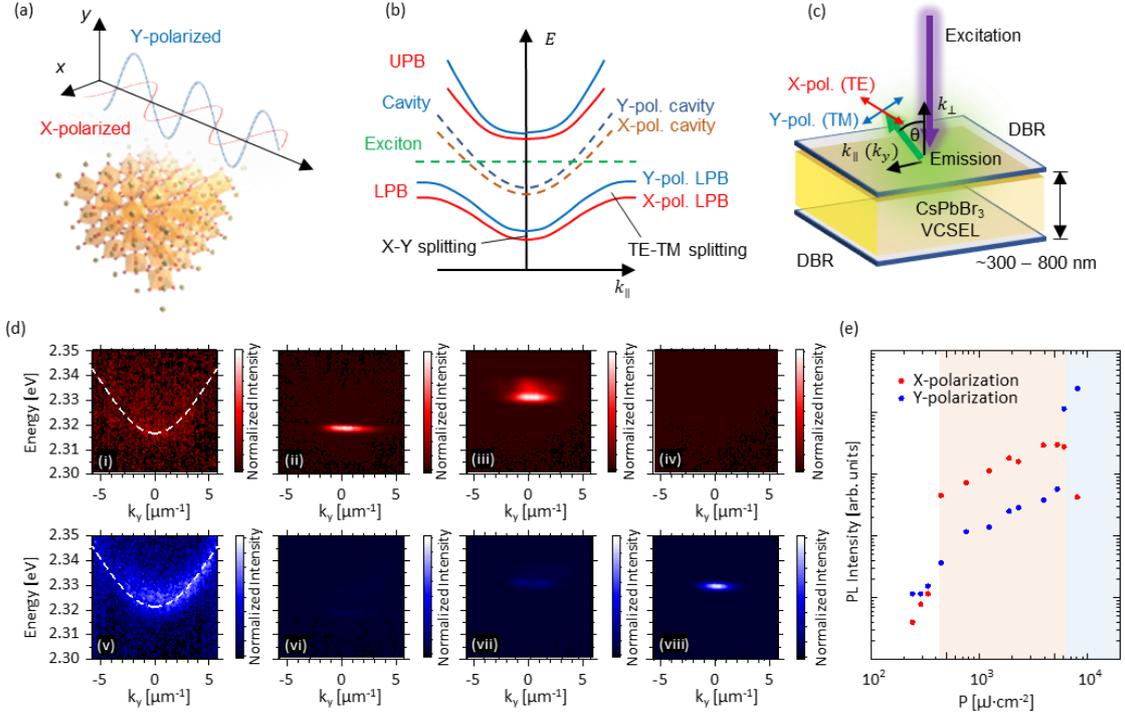

**Figure 1.** CsPbBr$_3$ microcavity with X-Y splitting. (a) A schematic of a CsPbBr$_3$ crystal and electromagnetic waves with orthogonal polarizations. (b) A schematic of dispersion curves with X-Y splitting and TE-TM splitting in LPB and UPB modes. (c) A schematic showing the configuration of angular-dependent PL measurement. (d) Colormaps of X-polarized [panels (i – iv)] and Y-polarized [panels (v – viii)] angle-resolved PL. Pump fluences $P$ are (i, v) 245, (ii, vi) 449, (iii, vii) 1260, and (iv, viii) 8170 μJ/cm$^2$. (e) $P$-dependence of PL signal intensity at the Γ point. Red and blue circles plot the data at X- and Y-polarizations, respectively.

We show angle-resolved PL results for a CsPbBr$_3$ microcavity fabricated with a solution-based crystallization method.[25,28–31] Figures S1, S2, and S3 in the Supplementary Information for the fabrication procedure, microscopic images of fabricated crystals, and optical properties of microcavity, respectively. Schematic diagram of the Fourier-space imaging setup is shown in Fig. S4 in the Supplementary Information. One can find details of the experimental methods in the Method section. In the angle-resolved PL measurements, X- and Y-polarizations correspond to TE and TM polarizations, respectively [see Fig. 1(c)]. Figure 1(d) shows colormaps of linearly polarized angle-resolved PL results obtained under the various pumping fluence $P$. The full dataset is shown in Fig. S5 in the Supplementary Information. At $P$ as low as ~245 µJ/cm$^2$, we find that the LPB modes at the both polarizations are in the strong coupling regime and show a parabolic dependence on the in-plane wavevector $k_y$ around zero [see panels (i) and (v) of Fig. 1(d)]. Rabi splitting energies are evaluated from the fitting analyses using a coupled oscillator model to be 62 meV and 45 meV for the X-polarized and Y-polarized LPB modes, respectively (see Supplementary text S1 and Fig. S6 in the Supplementary Information). The X-Y splitting energy at $k_y$ ~ 0 is ~6.1 meV at the lowest $P$ in this measurement. The mode number for the X- and Y-polarized modes is expected from the group index (~2.4) and the crystal thickness (~440 nm) to be 3.

In the X-polarization, the parabolic LPB mode disappears at $P$ ~449 µJ/cm$^2$, and instead, a highly directional PL signal appears at ~2.316 eV [see panel (ii) of Fig. 1(d)]. Furthermore, the emission intensity that reflects the polariton density shows a threshold behavior [see Fig. 1(e)], indicating the RT condensation of polariton particles. With increasing $P$, the energy of condensed state shows a blue shift [see panel (iii) of Fig. 1(d)] caused by polariton-polariton interaction. It should be noted that the energy of blue-shifted condensed state is ~2.330 eV, which already exceeds the energy of Y-polarized LPB mode at $k_y$ ~ 0 [~2.322 eV, see panel (v) of Fig. 1(d)].

A further increase in $P$ (> ~6150 µJ/cm$^2$) leads to the second threshold behavior for the Y-polarized LPB mode [see panel (viii) of Fig. 1(d)]. Interestingly, we find a drastic switching in the PL signal intensities between the two polarized states at the second threshold, where the Y-condensed state appears and the X-condensed state almost disappears [see panel (iv) of Fig. 1(d)], i.e. polarization switching. This phenomenon is observed for the first time in this study and is clearly different from a previous observation showing independent polariton lasing from orthogonally polarized LPB modes.[23]

To track detailed behaviors of the X- and Y-polarized states, we exhibit $P$-dependent PL spectra at $k_x = k_y \sim 0$ [see Fig. 2(a)]. We divide the experimental results into three pumping fluence ranges; namely, the first range is below the first threshold ($P < \sim 340$ μJ/cm$^2$), the second range is between the first threshold and the second threshold, and the third one is above the second threshold ($P > \sim 5330$ μJ/cm$^2$). In the range below the first threshold, the X-Y split two LPB modes are independently observed. This is seen as two polarization degree (PD) signals with the opposite signs, as shown in panels (i) – (iii) of Fig. 2(b). Here the PD is defined as

$$PD = \frac{I_X - I_Y}{I_X + I_Y}, \qquad (1)$$

where $I_X$ and $I_Y$ are PL intensities at the X- and Y-polarizations, respectively. The X-polarized signal shows an onset of condensation at $P$ just before the threshold.

In the second range, a PL signal showing the polariton condensation is observed at ~2.320 eV in the X polarization [see panel (iv) of Fig. 2(b)]. In addition to the conventional blue shift with the increased $P$, we find that the condensed mode is not perfectly polarized to the X direction but exhibits a small Y-polarized component. This fact is clearly found in normalized PL colormaps shown in Fig. S7 in the Supplementary information. More importantly, the blue-shifted condensed state exhibits an "energy hopping" from ~2.32 eV to ~2.33 eV at $P$ ~1260 μJ/cm$^2$ [see panels (v) and (vi) in Fig. 2(b)]. These results indicate the occurrence of X-Y polarization coupling in the polariton condensed states. This polarization coupling is made possible by the energetic resonance between the blue-shifted X-polarized condensed state and the Y-polarized LPB mode. It is worth noting that the coupling can be established even though the Y-polarized mode may be still incoherent. Meanwhile, non-zero interaction between the two modes is achieved by the orthorhombic crystal structure of CsPbBr$_3$ and/or another unexpected effect, which will be discussed later in detail, leading to the polarization hybridization of polariton condensed state. Note that our results do not show independent condensations of the X- and Y-polarized LPB modes because the energy of the two orthogonally polarized PL are the same within this pumping fluence range.

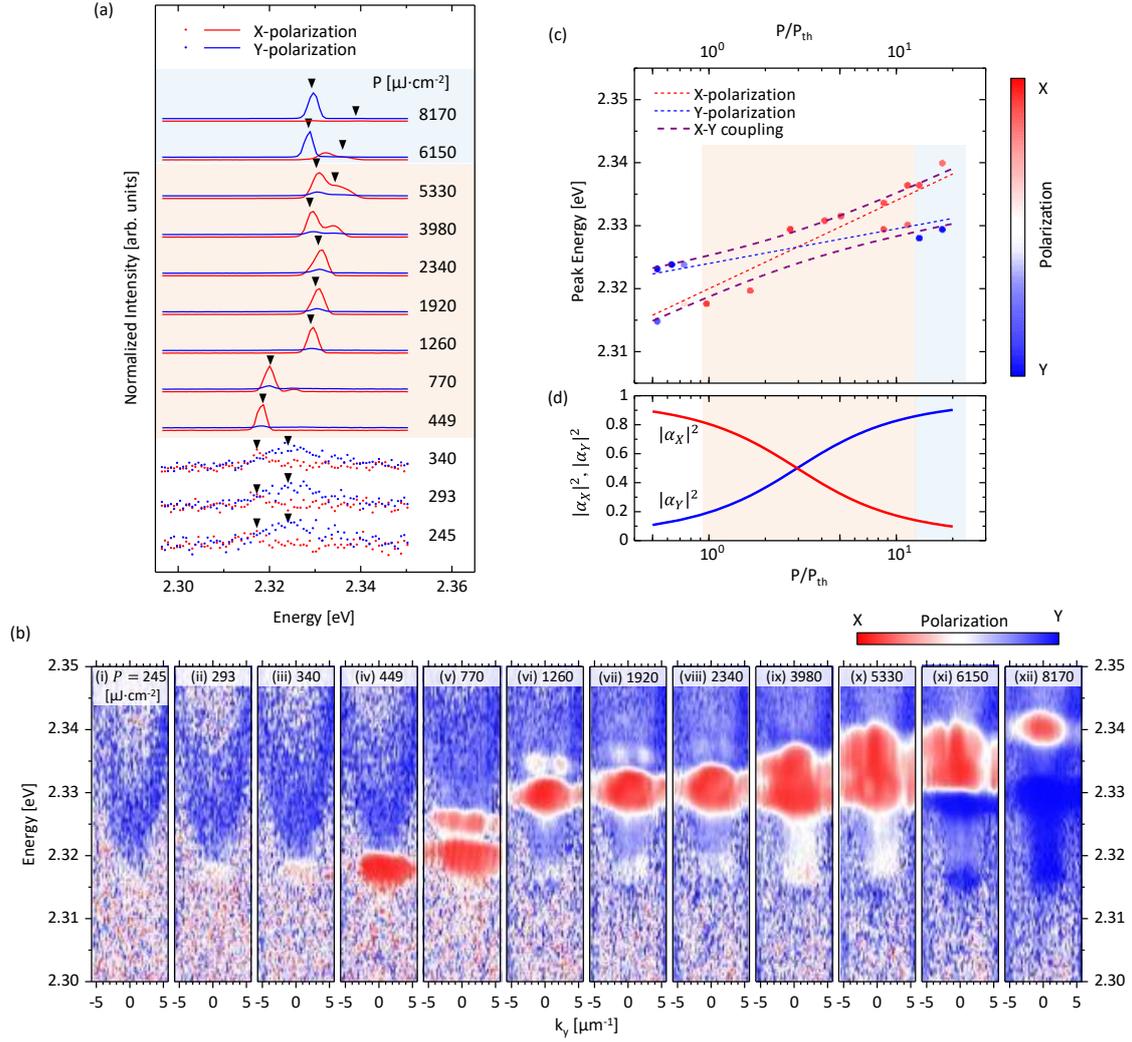

**Figure 2.** Polarization coupling of polariton condensed state in CsPbBr$_3$ microcavity. (a) $P$-dependent PL spectra at the Γ point. Red and blue curves show X- and Y- polarization, respectively. Orange and blue shaded region show the second and third ranges of $P$, which describe in the main text. No shaded region indicates the first range of $P$. (b) Colormaps of polarization degrees for polariton signals recorded at various $P$. (c) Peak energies of PL signals as functions of $P$. Colors of each plot shows the polarization degree of the signals. Dashed lines exhibit functions derived from a fitting analysis using equations (2) and (3). (d) $|\alpha_X|^2$ and $|\alpha_Y|^2$ for polarization coupled LPB mode calculated from the fitting results. This result exhibits one of the results for the lower mode. For the higher mode, curves for $|\alpha_X|^2$ and $|\alpha_Y|^2$ are simply exchanged.

In the third pumping fluence range ($P > \sim 6150$ μJ/cm$^2$), the X-polarized condensed state shifts to ~2.34 eV and is completely separated from the energy of Y-polarized state [see panels (xi and xii) of Fig. 2(b)]. Meanwhile, the polarization of condensed state at ~2.33 eV switches to Y-polarization, leading to the polarization switching.

We employ a simple model to formulate our results. The coupling between the X- and Y-polarized states is described well by a coupling model with a phenomenological Hamiltonian of

$$\begin{bmatrix} E_X & V \\ V & E_Y \end{bmatrix} \begin{bmatrix} \alpha_X \\ \alpha_Y \end{bmatrix} = E_c \begin{bmatrix} \alpha_X \\ \alpha_Y \end{bmatrix}. \quad (2)$$

Here, $E_X$ and $E_Y$ are energies of uncoupled X- and Y-polarized states, and $E_c$ is the eigenenergy of X-Y coupled state. We assume that the interaction term $V$ is constant against the polariton density. $|\alpha_X|^2$ and $|\alpha_Y|^2$ show the fractions of X- and Y-polarized states, respectively. In Gross-Pitaevskii equation, the blue shift of $E_X$ or $E_Y$ ($\Delta E$) is described to be proportional to the density of polariton particles, $N_{pol}$.[1,32–34] On the other hand, previous reports have shown that the experimentally observed $\Delta E$ follows a logarithmic function on $P$.[35,36] We employ the latter case and express $\Delta E$ as a logarithmic function on $P$.

$$\Delta E(P/P_{th}) = C \log(P/P_{th} - 1) \quad (3)$$

Here, $C$ is a constant and $P/P_{th}$ is the pump fluence normalized by that at the threshold.

Figure 2(c) compares the experimental result and the theoretical model shown above and reveals that the model can express the experimental result well. $V$ is estimated to be ~2.6 meV. In Fig. 2(d), we find that the two polarized states are hybridized well when $|\alpha_X|^2$ and $|\alpha_Y|^2$ are in a range of 0.2 – 0.8, which approximately corresponds to the second range on $P$. Interestingly, the condensed mode in this range is successfully "initialized" to the X polarization by the non-resonant pulse pumping. When $|\alpha_X|^2$ and $|\alpha_Y|^2$ are outside this range, the hybridization becomes small, and thus the independent polariton condensation appear in the third range of $P$.

Finally, we explain the origins of interaction between the X- and Y-polarizations. As shown in a recent paper,[37] the mode coupling equation at an energetic resonant condition between the orthogonally polarized cavity photon modes with the same number of antinodes is expressed as

$$E \begin{bmatrix} \alpha_X \\ \alpha_Y \end{bmatrix} = \left\{ \begin{bmatrix} E_X & 0 \\ 0 & E_Y \end{bmatrix} + \frac{(\hbar c)^2}{\varepsilon_{zz}} \begin{bmatrix} \varepsilon_{xx} k_x^2 + \tilde{\varepsilon}_{zz} k_y^2 & (\varepsilon_{yy} - \varepsilon_{zz}) k_y k_x \\ (\varepsilon_{xx} - \tilde{\varepsilon}_{zz}) k_y k_x & \varepsilon_{zz} k_x^2 + \varepsilon_{yy} k_y^2 \end{bmatrix} \right\} \begin{bmatrix} \alpha_X \\ \alpha_Y \end{bmatrix}. \quad (4)$$

Here $\varepsilon_{ij}$ denotes the dielectric tensor inside the cavity, and $\tilde{\varepsilon}_{zz} = \varepsilon_{zz} + \varepsilon_{xz}\varepsilon_{zx}/\varepsilon_{zz}$. Note that equation (4) is equivalent to equation (2), and we focus on the non-diagonal terms here. For CsPbBr$_3$ showing the orthorhombic crystal structure at RT, we thus obtain $\varepsilon_{xx} \neq \varepsilon_{yy} \neq \varepsilon_{zz}$. Furthermore, as the wavefunction of polariton condensed state is localized in the real space and shows a spread in the wavenumber space, we should consider that $k_y k_x$ is not perfectly zero. As a consequence, the off-diagonal terms, $(\varepsilon_{xx} - \tilde{\varepsilon}_{zz}) k_y k_x$ and $(\varepsilon_{yy} - \varepsilon_{zz}) k_y k_x$, are found to be non-zero in a CsPbBr$_3$ microcavity, showing that the polarization coupling is possible. Note that the off-diagonal terms become zero in the case of crystal showing a high symmetry (e.g. cubic) or in the case of different numbers of antinodes for the two polarized LPB modes. From these aspects, it is noteworthy that the CsPbBr$_3$ microcavity can fulfill these two requirements simultaneously owing to the 'very small' anisotropy. We know that the non-zero off-diagonal terms discussed here might be small values, but in a cavity with high quality factor, the overall hybridization degree is significantly enhanced in an optical cavity.

In addition to the non-zero off-diagonal terms discussed above, we consider another unexpected reason for the X-Y polarization coupling. Figure S8 in the Supplementally Information shows the angle-resolved PL results for another microcavity samples with different X-Y detuning. In these samples, we observe LPB modes with the X-Y splitting, but they do not show clean polarization features. This result implies that the crystal axis of CsPbBr$_3$ deviates from the normal direction of the microcavity. Even a microcavity showing an almost perfectly polarized LPB mode may have a slight misalignment of the crystal axis, leading to the nonzero off-diagonal terms in the permittivity tensor projected into the microcavity plane. Thus, crystal growth techniques is also an important point to manipulate $V$ strictly and to realize the practical application of RT polariton devices.

Conclusions

We realized a polarization superposition of RT polariton condensed states. CsPbBr$_3$ crystal, which has proven to be an excellent medium for the RT polariton condensation, exhibits an anisotropic feature suitable for the realization of polarization superposition. Specifically, the orthorhombic crystal structure of CsPbBr$_3$ can address simultaneously the two requirements: a small X-Y splitting and non-zero off-diagonal interaction terms. By exploiting the blue shift of polariton condensed state, we achieved a dynamical energy tuning between the two polarized modes, leading to the polarization superposition of RT polariton condensed state. The condensed state was found to be "initialized" to one of the polarization directions by non-resonant optical pulse pumping. As a result of polarization coupling, furthermore, we observed a polarization switching phenomenon.

The results shown here implies a great promise of polariton as RT quantum technologies. The polarization superposition is considerable progress toward the realization of RT polariton qubit. Future research should include demonstration of interactions between the polariton qubits and the realization of quantum logic gate operations. These challenges would open new perspectives on quantum applications of polaritons.

Methods

*Microcavity fabrication*

Distributed Bragg reflectors (DBRs) used in this study were multilayers of SiO$_2$ and TiO$_2$ on BK7 glass plates (area: $10 \times 10$ mm$^2$ and thickness 0.5 mm). The rf-magnetron sputtering method was employed for deposition of nine pairs of SiO$_2$ and TiO$_2$ layers. In this study, we used two types of DBRs. One of them had a reflection band (> 99 %) of ~450 – 550 nm and was used as the top mirror. The other had a reflection band of ~500 – 600 nm and was used as the bottom mirror.

To prepare precursor solutions of CsPbBr$_3$ perovskite, cesium bromide (CsBr) and lead(II) bromide (PbBr$_2$) were purchased from TCI. Dimethyl sulfoxide (DMSO) was purchased from Nacalai Tesque. We dissolve CsBr and PbBr$_2$ in DMSO in molar ratio 1:1 at a concentration of 0.5 M, stirred at 700 rpm with a screw tube and a stirring bar for ~1 hour at 50 °C.

CsPbBr3 microcavities were fabricated by a solution-processed crystallization method as shown in Fig. S1 in the Supplementary Information. First, two DBR mirrors mentioned above were washed with ultrasonic cleaning in ethanol and acetone for 10 minutes each and with UV ozone for 15 minutes. We casted the 10 μl of supersaturated precursor solution on a bottom DBR mirror. After that, the sample was capped with a top DBR mirror and then pinched with a double clip (19 mm). The crystal was grown at room temperature for one day. Next we injected 1 μl of precursor solution into the gap of DBR mirrors using a micropipette, from the three sides on the sample. Then the crystal was grown further for one day at room temperature. This procedure was repeated for 4~5 times without intervals. Finally, we injected a pure solvent into the gap and dissolved all crystals once. The sample was left in atmospheric condition for 3~4 days at 50 °C to re-grow a large-sized crystal again. After the growth, we removed the DMSO solvent using ethanol. Three days later, we removed the double clip. The thickness of crystal depended on the strength of the clipping and the amount of solution at the casting stage and was typically in a range of 300 – 700 nm.

*Characterizations*

All optical measurements were performed in atmospheric condition (~23 °C and ~40 %RH). We used a pulse laser (FTSS355-Q4, CryLas) for sample excitation. The emission spectra were measured with a CCD spectrometer (Kymera & Newton, Oxford ANDOR). The emission counts were recorded with time-integration of 1 s and accumulation of 10 times. In angle-resolved PL measurements, we used a Fourier space imaging setup (see Fig. S4 in the Supplementary Information) using an objective with NA ~ 0.5 (UPLFLN 20X, Olympus). The angle limits of angle resolved measurement is estimated from a relationship of $\theta_{max} = \sin^{-1}(\text{NA})$ to be 30 °. Polarization measurements were performed by using a Glan-Thompson prism as the polarizer.

Data availability

The datasets generated during and/or analysed during the current study are available from the corresponding author on reasonable request.

Acknowledgements

K.Y. acknowledges funding from Japan Society for the Promotion of Science, JSPS KAKENHI (Nos. 20KK0088, 22K18794, and 22H00215) and from JST CREST (JPMJCR02T4).

Conflict of interest

The authors declare no conflict of interests.

Author contributions

Y.M. and K.Y. conceived and planned the experiments. Y.M., T.I., and Y.U. performed he optimization of crystal growth and device fabrication. Y.M., T.I., and T.H. performed the construction of Fourier-space imaging setup and optical measurements. Y.M., S.T., and K.Y. deeply discussed the experimental results. Y.M. and K.Y. drafted the manuscript and complied figures, with discussion of results and feedback from all authors.